 \documentclass[acmsmall]{acmart}

\AtBeginDocument{%
  \providecommand\BibTeX{{%
    \normalfont B\kern-0.5em{\scshape i\kern-0.25em b}\kern-0.8em\TeX}}}

\setcopyright{acmcopyright}
\copyrightyear{2025}
\acmYear{2025}
\acmDOI{XXXXXXX.XXXXXXX}

\acmConference[CSCW '25]{Make sure to enter the correct
  conference title from your rights confirmation emai}{October 18--22,
  2025}{Bergen, Norway}
\acmPrice{15.00}
\acmISBN{978-1-4503-XXXX-X/18/06}

\usepackage[flushleft]{threeparttable} 
\usepackage{makecell} 
\usepackage{soul}

\begin{document}

\title{SnuggleSense: Empowering Online Harm Survivors Through a Structured Sensemaking Process}


\author{Sijia Xiao}
\affiliation{%
  \institution{Carnegie Mellon University}
  \country{USA}}
\email{xiaosijia@cmu.edu}

\author{Haodi Zou}
\affiliation{%
  \institution{University of California, San Diego}
  \country{USA}}
\email{h4zou@ucsd.edu}

\author{Amy Mathews}
\affiliation{%
  \institution{University of California, Berkeley}
  \country{USA}}
\email{amymathews@berkeley.edu}

\author{Jingshu Rui}
\affiliation{%
  \institution{University of California, Berkeley}
  \country{USA}}
\email{jingshu_rui@berkeley.edu}

\author{Coye Cheshire}
\affiliation{%
  \institution{University of California, Berkeley}
  \country{USA}}
\email{coye@berkeley.edu}

\author{Niloufar Salehi}
\affiliation{%
  \institution{University of California, Berkeley}
  \country{USA}}
\email{nsalehi@berkeley.edu}

\renewcommand{\shortauthors}{Xiao et al.}
\renewcommand{\shorttitle}{SnuggleSense}

\begin{abstract}
Online interpersonal harm, such as cyberbullying and sexual harassment, remains a pervasive issue on social media platforms. Traditional approaches, primarily content moderation, often overlook survivors' needs and agency. We introduce SnuggleSense, a system that empowers survivors through structured sensemaking.
Inspired by restorative justice practices, SnuggleSense guides survivors through reflective questions, offers personalized recommendations from similar survivors, and visualizes plans using interactive sticky notes.
A controlled experiment demonstrates that SnuggleSense significantly enhances sensemaking compared to an unstructured process of making sense of the harm.
We argue that SnuggleSense fosters community awareness, cultivates a supportive survivor network, and promotes a restorative justice-oriented approach toward restoration and healing. We also discuss design insights, such as tailoring informational support and providing guidance while preserving survivors' agency.
\end{abstract}

\ccsdesc[500]{Computer systems organization~Embedded systems}
\ccsdesc[300]{Computer systems organization~Redundancy}
\ccsdesc{Computer systems organization~Robotics}
\ccsdesc[100]{Networks~Network reliability}

\keywords{datasets, neural networks, gaze detection, text tagging}



\maketitle

\section{Introduction}
Interpersonal harm, such as cyberbullying and sexual harassment, is a pressing issue on social media platforms  \cite{duggan_online_2017, vogels2021state}. 
Online platforms primarily address these types of harm through content moderation, which focuses on punishing perpetrators through actions such as content removal or account banning. Survivors are often left out of decision-making in this perpetrator-centered framework. Research has found that survivors have unmet needs, including seeking advice, obtaining emotional support, and receiving acknowledgment and an apology from the person who caused the harm \cite{xiao2022sensemaking, schoenebeck2021drawing, musgrave2022experiences}. 

Given the growing scale and ramifications of online harm \cite{duggan_online_2017, vogels2021state}, empowering survivors is a matter of societal and ethical urgency. 
In recent years, there has been growing interest within the fields of CSCW and HCI to adopt a survivor-centered approach by prioritizing the needs and agency of survivors and by providing tools and resources to support their harm resolution \cite{sultana2022shishushurokkha, schoenebeck2021drawing, musgrave2022experiences, goyal2022you, dimond2013hollaback}. We build on this line of work by focusing on a critical but underexplored aspect of survivor empowerment: \textit{sensemaking}. Sensemaking is the process through which survivors gather information to understand the harm, recognize the resources available to them, and develop a plan of action to address their needs \cite{weick1995sensemaking}. Research has found that it can be challenging for survivors to make sense of what they need and the actions to meet those needs within a perpetrator-centered content moderation process \cite{xiao2023addressing}. Survivors face challenges when seeking support in the sensemaking of harm, such as difficulty assessing the impact and severity of the harm \cite{andalibi2021sensemaking, goyal2022you} and uncertainty about where to seek help \cite{xiao2022sensemaking, to2020they}.

While existing research has explored tools that mobilize survivors to take action such as documenting evidence \cite{sultana_unmochon_2021, goyal2022you} or mobilizing friend groups for support \cite{maharSquadboxToolCombat2018}, fewer systems explicitly focus on assisting survivors in the process of sensemaking. However, as an early stage in harm resolution \cite{xiao2022sensemaking}, a lack of clarity in sensemaking may hinder survivors' ability to determine what actions to take and whom to turn to for support in addressing their harm. In this paper, we introduce \textit{SnuggleSense}, a system designed to empower survivors through a structured sensemaking process. After survivors experience harm, SnuggleSense facilitates a process for them to understand the harm and develop a plan of action, especially in situations where immediate support is not available, or survivors are hesitant to reach out due to the fear of secondary harm. 

To achieve this, we draw inspiration from a survivor-centered justice framework - restorative justice. Restorative justice is both a practice and philosophy of justice that prioritizes survivors' agency and needs in addressing harm. In recent years, researchers from CHI, CSCW and other related fields have applied restorative justice to comprehend online harm and provide support to survivors \cite{schoenebeck2021drawing, kou2021punishment, hughes2020keeper, xiao2022sensemaking, ngoc2025design}, extending its application beyond traditional offline settings such as the criminal justice system, schools, and workplaces \cite{van2016overview, wood2016four}. SnuggleSense follows this line of work and explores how we can apply restorative justice to expand the toolkit available to survivors for making sense of online harm.

SnuggleSense draws inspiration from two restorative justice practices: pre-conference and circles. First, SnuggleSense guides survivors through reflective questions inspired by the pre-conference process, where survivors work with a trained facilitator to process harm, identify needs, and develop an action plan \cite{zehr2015little, xiao2022sensemaking}. Second, SnuggleSense incorporates the social support aspect of circles by offering suggested actions from other survivors who have undergone similar experiences of harm. These elements are integrated into a design process facilitated by interactive digital sticky notes. The final outcome of the sensemaking process is a series of sticky notes arranged on a timeline, representing a step-by-step plan for addressing the harm in chronological order.

We compared how SnuggleSense facilitated survivors' sensemaking process to how they typically make sense of harm within the content moderation framework. We conducted a within-subject, controlled experiment where survivors developed an action plan for the harm they experienced using either SnuggleSense or by writing out the plan themselves. Our results indicate that participants found SnuggleSense significantly more effective in facilitating their sensemaking of harm. It increases survivors' awareness of available resources and community, offering a pathway for addressing harm that emphasizes healing and restoration. We discuss the implications of SnuggleSense’s design, including tailoring support to individual survivors, fostering a support community, ensuring safeguards for survivors as the system scales, and facilitating meaningful action following the sensemaking process.

\section{Related Work}
\subsection{Online Interpersonal Harm, Content Moderation and Survivor-Centered Approaches}
Various forms of harm can occur on the internet, such as breach of privacy \cite{pesce2012privacy}, misinformation and disinformation \cite{xiao2021sensemaking, starbird2019disinformation}, digital self-harm \cite{pater2017defining}, hate speech \cite{das2020hate}, or bullying and harassment \cite{vogels2021state}. In our research, the harm we investigate falls under the definition of interpersonal harm, which is defined as offensive behavior directed towards individuals by other individuals \cite{zeelenberg2008role}. Online interpersonal harm disproportionately affects certain demographic groups, including young adults, women, and Black and Hispanic individuals \cite{vogels2021state}. It can have a wide range of negative consequences, including psychological distress \cite{vitak2017identifying, blackwell2019harassment} and a decrease in engagement in both online and offline activities \cite{fox2017women, celuch2021longitudinal}. These negative effects can have a long-lasting impact on the affected individuals and can make them more vulnerable to future harm \cite{celuch2021longitudinal, schoenebeck2021drawing}. 

Online platforms currently address interpersonal harm through content moderation, which involves taking actions to penalize perpetrators of harm based on the severity of their offense \cite{roberts2019behind}. These measures can include removing offensive content, banning or warning users \cite{gillespie2018custodians}.  Researchers have proposed ways to improve content moderation such as increasing transparency \cite{jhaver2019does}, involving users in moderation efforts \cite{vaccaro2021contestability}, or utilizing algorithms and bots to moderate at scale \cite{chandrasekharan2019crossmod, seering2019moderator}. However, despite these efforts to improve content moderation, the severity of interpersonal harm such as harassment has continued to increase over time \cite{vogels2021state, duggan_online_2017}. 

In recent years, a growing body of research has embraced a survivor-centered approach to comprehending the experiences and needs of individuals facing online harm \cite{sultana2022shishushurokkha, schoenebeck2021drawing, musgrave2022experiences, goyal2022you, dimond2013hollaback}. Studies have revealed that survivors possess needs that content moderation alone cannot adequately address, including the need to make sense of their experiences, receive emotional support, and contribute to the transformation of the online environment \cite{xiao2022sensemaking, to2020they}. There exists an urgent requirement to provide guidance, support, and resources to assist survivors in fulfilling these needs \cite{thomasItCommonPart2022, xiao2022sensemaking}. Researchers have studied survivors' experiences and challenges in addressing harm, exploring both collective sensemaking and empowerment \cite{musgrave2022experiences, blackwell_classification_2017}, as well as individualized endeavors and needs \cite{im2022women, schoenebeck2021drawing, blackwell_when_2018}. 

Researchers have also studied how social-computing platforms can support people who experience harm. These studies often feature online support communities that focus on people who experience harm in an offline context, such as pregnancy loss \cite{barta2023similar}, depression \cite{andalibi2017sensitive}, or the COVID-19 pandemic \cite{10.1145/3617654}. Most of these communities are on social media platforms and thus adopt the platform's content moderation to regulate the content \cite{10.1145/3617654, andalibi2017sensitive, dimond2013hollaback}. Despite the scale of online harm and the unique challenges online survivors face, there are rarely communities that focus on the experiences of online harm survivors. In recent years, researchers have developed platforms or tools specifically designed to support online harm survivors in addressing their needs. Some tools try to provide social support by asking survivors' friends to review their messages \cite{maharSquadboxToolCombat2018} or by allowing survivors to share their experiences and receive advice and help for reporting online harassers \cite{blackwell_classification_2017}. Others have built tools to help online harm survivors document evidence of harassment \cite{sultana_unmochon_2021, goyal2022you}. While these tools often aim to mobilize actions, SnuggleSense adds to this body of work by facilitating online harm survivors in their sensemaking process.


\subsection{Restorative Justice}
Content moderation mechanisms are based on principles of punitive justice, which view harm as a violation of rules and aim to punish perpetrators in proportion to their offense \cite{garland1993punishment}. In contrast, restorative justice views harm as a breach of interpersonal relationships and aims to promote healing and restoration for all parties involved \cite{zehr2015little}. This approach focuses on providing support and healing to survivors, helping perpetrators understand the harm they have caused, and engaging community members in addressing harm collectively \cite{mccold2000toward}.

Restorative justice tackles harm through communication and collaboration among all parties involved, aiming to achieve consensus on an action plan and subsequent implementation of the plan \cite{zehr2015little}. Survivors, as the subjects who receive harm, play a crucial role in the restorative justice process. Empowerment of survivors is a central element of restorative justice, both as a procedural safeguard and a criterion for success  \cite{aertsen2011restorative}. The restorative justice approach emphasizes the importance of returning agency and power to survivors, who have often been disempowered by the traditional criminal justice system \cite{aertsen2011restorative}. This can be achieved through various actions, such as emotional validation and support from restorative justice facilitators and community members, and apologies from perpetrators \cite{daly2003restorative}. Our research aligns with the values and practices of restorative justice and aims to empower survivors by adopting its principles and practices.

Specifically, restorative justice practices encompass individual and collective sensemaking processes before the implementation of action plans. In a restorative justice pre-conference \cite{pranis2015little}, a facilitator guides survivors in reflecting on the harm and articulating their needs for addressing it. The subsequent process, known as circles, involves collective sensemaking of harm and the consensus-building on an action plan with other pertinent stakeholders, including perpetrators, community members, and family and friends \cite{johnstone2013handbook}. Our research draws inspiration from the pre-conference and circle processes as mechanisms for survivors to make sense of the harm they have experienced.

Restorative justice has been successfully applied in offline scenarios to address harm such as schools, workplaces and the criminal justice system \cite{van2016overview, wood2016four}. In recent years, researchers have begun to explore its potential for addressing online harm. Studies have examined survivors' needs and preferences for restorative justice actions in addressing online harm \cite{schoenebeck2021drawing, kou2021punishment}. Researchers have also designed tools that incorporate restorative justice practices. Xiao et al. has developed an online design activity inspired by pre-conference to identify survivors' needs in addressing online harm \cite{xiao2022sensemaking}. Doan et al. developed ApoloBot, a Discord bot to facilitate apologies when harm occurs in online communities inspired by restorative justice \cite{ngoc2025design}. Hughes and Roy developed Keeper, a tool to facilitate restorative justice discussions online \cite{hughes2020keeper}. Our tool joins this line of work by bringing restorative justice practices into the context of helping online harm survivors with sensemaking.

\section{System Design}
\begin{figure*}
    \centering
    \includegraphics[width=\textwidth]{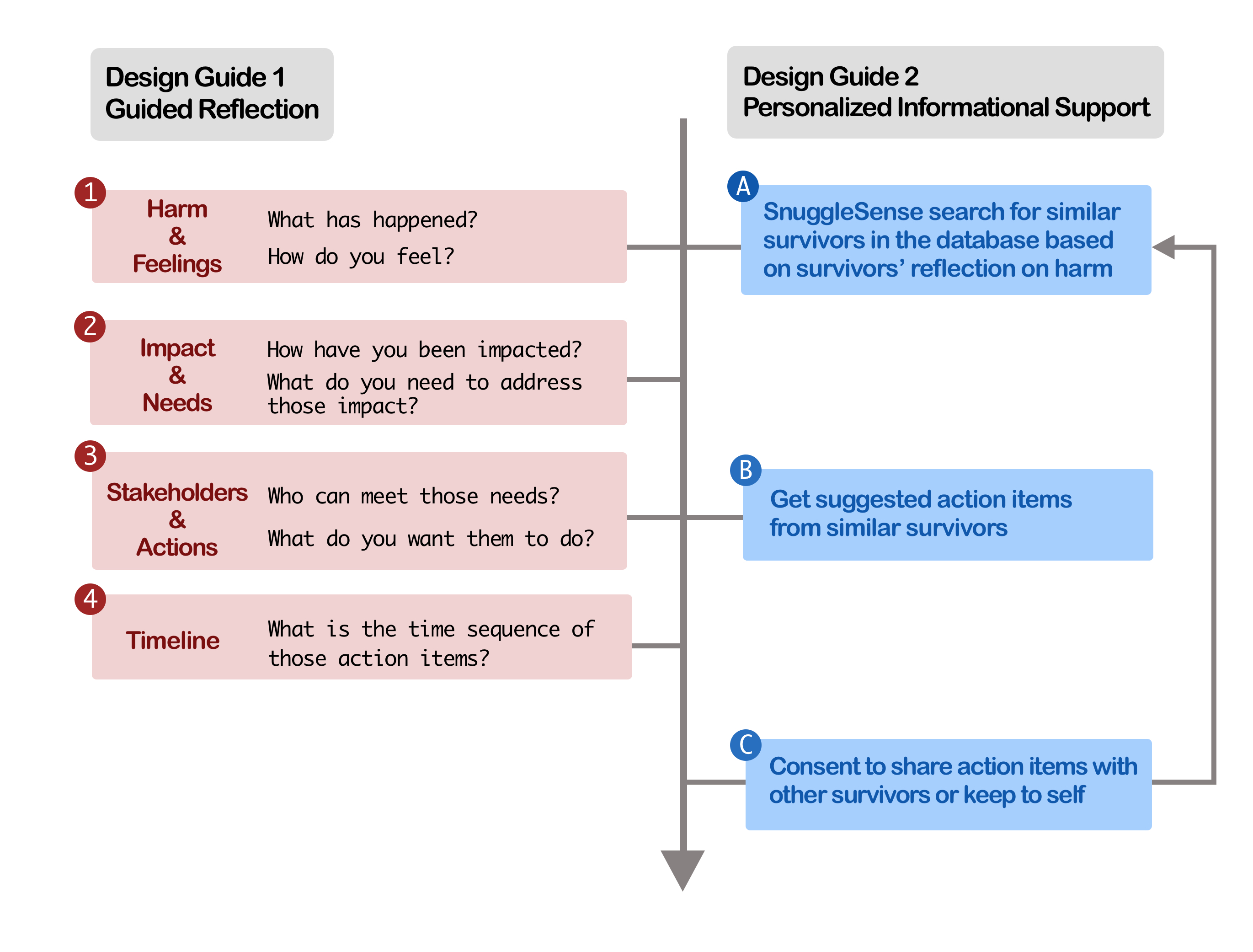}
    \caption{
    The Guided Reflection Process and Personalized Informational Support in SnuggleSense. SnuggleSense guides survivors' sensemaking process through a series of reflective questions inspired by restorative justice pre-conference. The questions prompt survivors to reflect on their experiences of harm, their feelings, the impact of the harm, their needs, and action plans to address those needs (steps 1-4). SnuggleSense also supports survivors' sensemaking process by providing them with personalized information. Based on each survivor's answer to the reflective questions, SnuggleSense searches for similar survivors in the database (step A) and recommends action items from similar survivors (step B). If consent is given (step C), survivors' action plans are incorporated into the database for making future suggestions.
    }
    \label{figdesign}
    \Description{
    }
\end{figure*}

\begin{figure*}
    \centering
    \includegraphics[width=\textwidth]{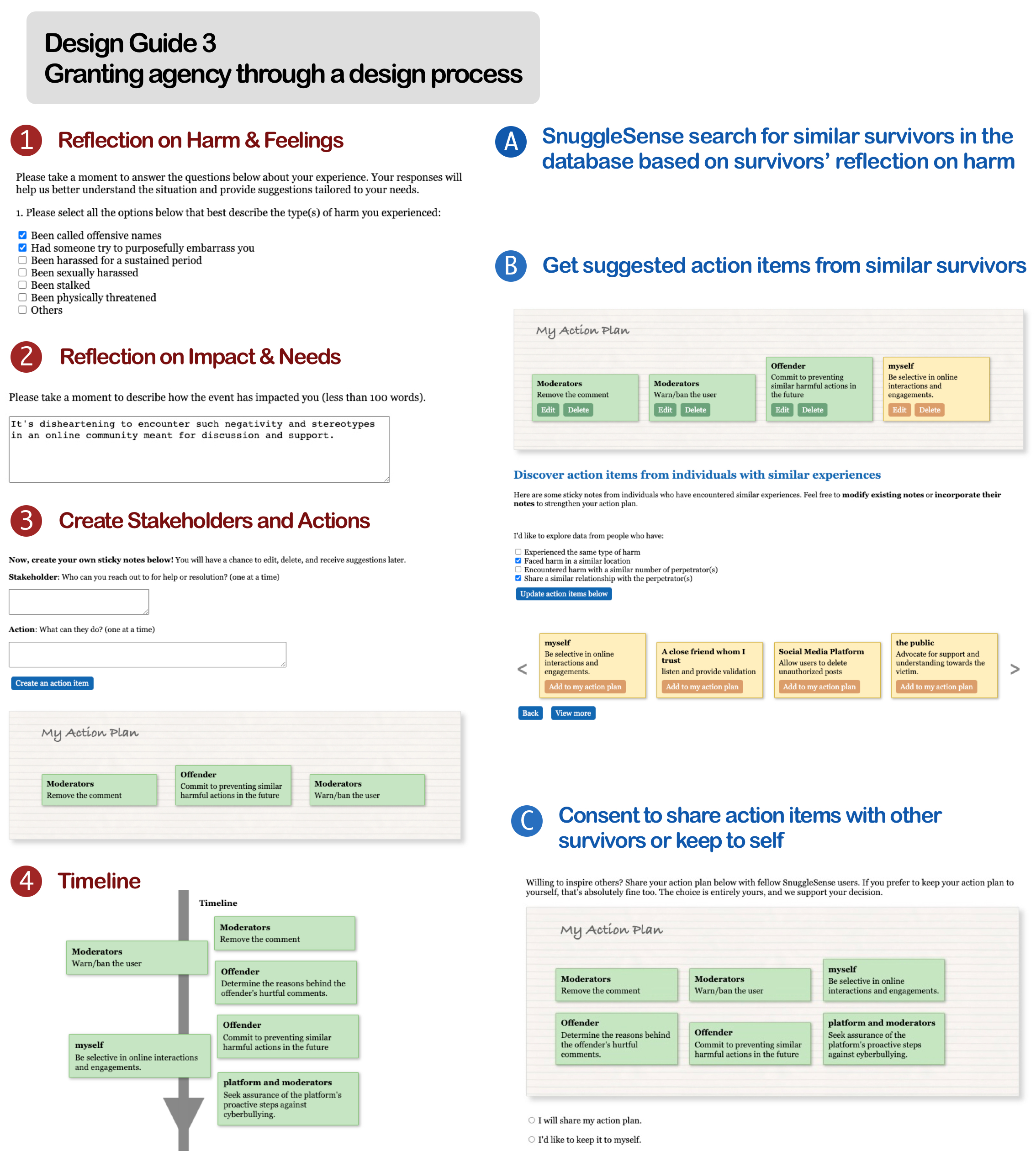}
    \caption{
    SnuggleSense Grants Agency Through a Design Process. The step number of the graph corresponds with Figure \ref{figdesign}. SnuggleSense grants survivors agency through interactive sticky notes and a visual timeline for their action plans.   Participants use these features to generate their action items (step 3), include suggested actions (step B), and visualize their plans on a timeline (step 4). These design activities serve to encourage survivors to exercise agency and creativity in exploring diverse ways to address harm and meet their unique needs. The screenshots in this figure illustrate the essential components of the system but do not encompass the entire interface. 
        }
    \label{figdesign3}
    \Description{
    }
\end{figure*}

The design of SnuggleSense is motivated by the lack of tools that support survivors' sensemaking needs and is inspired by restorative justice. In this section, we first outline the design principles of SnuggleSense, drawing from prior research on online harm survivors' needs and restorative justice practices. We then describe how these principles guided our implementation of SnuggleSense. Finally, we reflect on our positionality and how it informs the system design.

\subsection{Design Guides}
\subsubsection{Design Guide 1: Guided Reflection}
Our first design guide is to provide survivors with guidance in their sensemaking of harm. When content moderation is the primary approach to addressing harm, survivors may not be encouraged to consider their role in addressing it and often struggle to envision solutions beyond this framework \cite{xiao2023addressing}. Survivors frequently need support when trying to make sense of the harm they've experienced \cite{xiao2022sensemaking, maengDesigningEvaluatingChatbot2022, thomasItCommonPart2022}.

SnuggleSense guides survivors through reflection questions inspired by restorative justice pre-conferencing, a step where a facilitator works with survivors to understand the harm and develop an action plan before engaging other stakeholders to reach consensus \cite{zehr2015little}. Salehi has drawn a comparison between the core questions the content moderation process and the restorative justice process ask \cite{salehi2020no}. In a content moderation process, these questions revolve around identifying reported content, determining its compliance with established rules, and deciding on appropriate actions such as removal, demotion, flagging, or ignoring. In contrast, the restorative justice process centers on different inquiries: Who has suffered harm? What are their needs? Whose responsibility is it to meet those needs?

SnuggleSense adheres to the questions above through the reflection process (Figure \ref{figdesign}, step 1-4). In this process, survivors first engage in introspection, reflecting on the harm and the emotions it has created within them (Figure \ref{figdesign}, step 1). Storytelling has a critical role in survivors' sensemaking, allowing them to recall and reconstruct their experiences from their unique perspective, thereby centering their emotions and experiences in the process \cite{zehr2015little}. Furthermore, storytelling can help resurface the details of the harm to aid in further reflections.

Once survivors have reflected on the harm and their emotional responses, SnuggleSense aids them in identifying the impacts of the harm and their associated needs (Figure \ref{figdesign}, step 2). The restorative justice approach maintains that harm gives rise to impacts, and these impacts inform the needs of survivors \cite{salehi2020no}. Survivors often possess needs that conventional justice processes, such as punitive online content moderation, fail to address. These needs encompass elements like truth-telling, restoration, emotional support, and validation \cite{zehr2015little}. By encouraging reflection on impacts and needs before specifying concrete actions, this process enables survivors to conceive a wider range of potential strategies for addressing harm that extend beyond conventional content moderation approaches.

After the reflection on impacts and needs, SnuggleSense guides survivors in the formulation of an action plan (Figure \ref{figdesign}, steps 3-4). SnuggleSense's action plan comprises tasks assigned to various stakeholders (e.g., moderators, bystanders, family and friends). Restorative justice views the process of addressing harm as inherently multi-stakeholder, positioning survivors and perpetrators within their communities and recognizing the involvement of community members as vital contributors to the resolution process \cite{zehr2015little}. Research has also highlighted the complexity of addressing online harm, often necessitating the coordination of multiple stakeholders both online and offline \cite{freedUnderstandingDigitalSafetyExperiences2023, xiao2022sensemaking}. Following the online pre-conference procedure outlined by Xiao et al. \cite{xiao2022sensemaking}, SnuggleSense breaks down the creation of an action plan into three parts: identification of stakeholders who bear responsibility or can offer assistance to survivors (Figure \ref{figdesign}, step 3), identification of the actions these stakeholders can undertake to address the harm (Figure \ref{figdesign}, step 3), and organization of these actions in a chronological sequence (Figure \ref{figdesign}, step 4). 

\subsubsection{Design Guide 2: Informational Support from Survivors with Similar Experiences}
SnuggleSense provides a unique online support system for survivors who may not have immediate human assistance after experiencing harm. It fosters a virtual community by storing and sharing action plans from survivors who consent to share their experiences with others. The design of informational support in SnuggleSense is inspired by restorative justice circles \cite{zehr2015little}. In these circles, survivors, along with other community members affected by the harm — including perpetrators, family, and friends — convene after the pre-conference to collaboratively formulate concrete action plans. Participation is voluntary, contingent upon resource availability and the willingness and commitment of the involved parties.

SnuggleSense aligns with this approach in a virtual setting, providing individual survivors with a platform to engage with and support one another. When immediate human support is not often available for survivors in their circumstances, how can we design an online system to connect survivors with the support they need? The system groups survivors with similar experiences together and shares their action items as suggestions. After a new user enters information about their specific harm case, SnuggleSense pairs them with survivors who have encountered similar experiences (Figure \ref{figdesign}, step A). As users input their own action items, SnuggleSense presents them with action items from peers who have faced comparable situations (Figure \ref{figdesign}, step B). Upon finishing their action plan, users have the option to choose whether to share their data with other survivors or keep it private (Figure \ref{figdesign}, step C). This repository of action items becomes a tool for mutual help and understanding among users. To safeguard privacy, survivors are afforded the opportunity to review their data before deciding whether to share it or retain it confidentially.


\subsubsection{Design Guide 3: Granting Agency Through a Design Process}
SnuggleSense leverages spatial reasoning to support survivors, inviting them to present their action plan through interactive digital sticky notes and a visual timeline (Figure \ref{figdesign3}). Our approach to utilizing design to facilitate survivors' sensemaking draws inspiration from speculative design. According to Wakkary et al., design serves as a catalyst for exploring alternatives and redistributes the power of interpretation to the users \cite{wakkary2015material}. They contend that design possesses the capacity to act as a bridge, connecting our present reality with an imagined, critically transformed perspective of our world. Moreover, Gerber underscores the notion that design artifacts can function as instruments for actualizing users' visions and igniting discussions and creativity around these concepts \cite{gerber2018participatory}.

While addressing harm experienced by survivors differs from speculative design's focus on hypothetical scenarios, this design approach prompts survivors to contemplate ideals that transcend the constraints of the existing system. By drawing inspiration from speculative design, our aim is to use sticky note design activities to encourage survivors to exercise their agency and nurture a creative mindset, allowing them to explore a wide spectrum of approaches to addressing harm and meeting their unique needs.

\subsection{Implementation of SnuggleSense}
SnuggleSense is a web-based platform developed using a front-end stack that includes JavaScript, D3, jQuery, HTML, and CSS. On the back end, it is implemented in Python, leveraging the Flask framework, and is hosted on the Google Cloud Platform for data storage. The development of SnuggleSense followed an iterative design process, involving pilot testing and user feedback to refine and enhance its features. Next, we provide a detailed description of the SnuggleSense implementation.

\subsubsection{Reflection: Harm, Feelings, Impact, Needs} \label{reflection_of_harm}
In the initial phase of SnuggleSense, survivors are prompted to engage in self-reflection by documenting the harm they have experienced (Figure \ref{figdesign3}, steps 1). Initially, survivors provide a brief description of their experience in a text box. To facilitate this process, we have also included a set of multiple-choice questions aimed at encouraging survivors to consider various aspects of the harm they have endured. These multiple-choice questions not only stimulate survivors to examine their experiences from different perspectives but also serve as input for generating personalized recommendations at a later stage (Figure \ref{figdesign3}, step A). These questions cover four dimensions of the harm experiences that are relevant to their needs for addressing harm, including the nature of the harm, the location where it occurred, the number of individuals involved, and their relationship to the survivor. We selected these four dimensions of harm experiences based on pilot testing to determine which aspects participants found most relevant for identifying their needs. Following this, participants further reflect on the impact of the harm and their needs for addressing it by writing in text boxes (Figure \ref{figdesign3}, step 2).

\subsubsection{Create Action Items for Stakeholders}
Following the reflective phase, SnuggleSense guides survivors in the creation of an action plan (Figure \ref{figdesign3}, step 3). We employ sticky notes to represent individual action items. Initially, we provide a sample action plan with example actions. Subsequently, we prompt survivors to compose their own action item, comprising a specific stakeholder and the corresponding action aimed at addressing the harm they have experienced.

\subsubsection{Receive Recommendations from Survivors with Similar Experiences}
After survivors have drafted their action plans, SnuggleSense offers support by presenting action item suggestions from other survivors who have encountered similar experiences (Figure \ref{figdesign3}, step B). Four suggestion sticky notes are initially presented to users, with the option to access more suggestions if desired. Users can integrate these suggestions into their existing action plans by clicking on the "Add to My Action Plan" button on the suggested sticky notes.


SnuggleSense offers relevant suggestions by grouping survivors with similar harm experiences. SnuggleSense calculates the similarity between the current user and existing users in the database based on multiple-choice questions they have answered about the context of harm (Figure \ref{figdesign3}, step A). In the database, a similarity score $S \in [0, 1]$ is stored for each pair of users. The similarity score is calculated as follows: For each multiple-choice question with $n$ options, if both survivors either selected or did not select an option, $\frac{1}{n}$ is added to the similarity score. The total similarity score $S_{ij}$ between two survivors $i$ and $j$ is the sum of the individual scores across all questions:

$$
S_{ij} = \sum_{k=1}^{m} \frac{1}{n_k} \cdot I_{ik,jk}
$$

\begin{scriptsize}
In the equation, $m$ is the number of questions, $n_k$ is the number of options for question $k$, and $I_{ik,jk}$ is an indicator function that equals $1$ if both survivors $i$ and $j$ selected (or did not select) the same option for question $k$, and $0$ otherwise.
\end{scriptsize}

For each user, we identify three survivors with the highest similarity scores from the database and recommend their action items in a randomized order. Additionally, we provide users with four selection boxes representing different aspects of harm, allowing them to choose the most relevant suggestions based on their priorities. Once users finish drafting their action plans, we record the harm experiences and action plans they consent to share and store them in the database for future recommendations (Figure \ref{figdesign3}, step C).


\subsubsection{Organizing Action Items Chronologically}
Subsequently, SnuggleSense prompts survivors to organize their action items in a chronological sequence (Figure \ref{figdesign3}, step 4). Building on the research by Wong and Nguyen \cite{wongTimelinesWorldBuildingActivity2021} and Xiao et al. \cite{xiao2022sensemaking}, this step aims to help survivors visualize their action plans by listing the tasks in the order they intend to carry them out. Survivors have the flexibility to add additional action items as they construct their timelines.

\subsubsection{Sharing Action Plans with the Community}
In the final phase, SnuggleSense presents the completed action plan to survivors and inquires whether they would like to share it with fellow users of the system (Figure \ref{figdesign3}, step C). This feature encourages survivors to engage with a community of peers who can provide valuable insights and support in relation to their action plans with their consent. Participants can also choose to keep the action plan to themselves.

\subsection{Positionality Statement} \label{positionality}
The authors of this paper have expertise in the fields of online harm and content moderation, with some having personal experiences with online harm. We reside in a society where punitive justice is the prevailing method for addressing harm, yet we acknowledge the merits of restorative justice, which emphasizes healing and restoration. The lead author has received training in restorative justice facilitation and has directly assisted online harm survivors using this framework. These experiences underscore our commitment to a survivor-centered approach in addressing online harm and exploring alternative approaches beyond the conventional punitive model.

While we draw inspiration from restorative justice with its survivor-centered nature and successful offline practice in helping survivors' sensemaking, our intent is not to advocate for it as the exclusive or prioritized way to address harm. Restorative justice's applicability is context-specific and varies depending on the nature of harm and individual circumstances \cite{zehr2015little}. We recognize the potential of different justice models. Traditional content moderation, as a punitive approach, has demonstrated effectiveness in stopping the continuation of harm and reducing re-offense \cite{gillespie2018custodians, jhaver2019does}. Addressing systemic issues such as sexism or discrimination requires a transformative approach aimed at rectifying the underlying structural problems \cite{evansStructuralViolenceSocioeconomic2016}. Rather than advocating for a specific approach, our objective is to explore ways to empower survivors by drawing from alternative methods of addressing harm that are not traditionally adopted, thereby expanding the toolkit available to online harm survivors in addressing the harm they experience.

\section{Evaluation}
SnuggleSense aims to empower online harm survivors by providing a structured sensemaking process to enhance how they make sense of harm within the content moderation framework. To evaluate the system’s effectiveness, we conducted a controlled experiment comparing survivors’ action plans created with SnuggleSense to those developed through an unstructured sensemaking process, which reflects current approaches to making sense of harm

\subsection{Controlled Experiment: A Comparison Between Writing Text and Using SnuggleSense}
The experiment task was to produce an action plan for a harm case the participant experienced.
We employed a within-subject design \cite{dean1999design}, allowing participants to compare two sensemaking approaches: an "Unstructured" condition and a "Structured" condition.

In the Unstructured condition, participants were asked to develop an action plan by directly writing a sequence of action items, each specifying a stakeholder and their corresponding actions to address the harm. This written approach served as the control condition, simulating the natural progression of an unguided sensemaking process. In the Structured condition, participants were asked to use SnuggleSense’s guided sensemaking process to create an action plan, also structured by stakeholders and actions. Here, the action plan was presented on a visual timeline that utilized SnuggleSense’s digital sticky notes to organize each item chronologically. Participants were allocated 15 minutes to complete the action plan in each condition.

We opted for a within-subject design rather than a between-subject design \cite{dean1999design}. Our preliminary testing showed that participants often found it challenging to assess the effectiveness of their sensemaking or their sense of empowerment without being aware of alternative methods. A within-subject experiment allowed us to directly compare how participants' sensemaking experiences differed when applied to the same harm scenario \cite{charness2012experimental}. To minimize potential priming effects, we presented the two conditions to participants in a randomized order \cite{molden2014understanding}.

We recruited individuals who had encountered harm in the past six months and asked them to reflect on an instance of harm that occurred within the designated timeframe set for this experiment. In our preliminary testing, we observed that individuals who had experienced harm a considerable time ago often had already engaged in substantial sensemaking and had developed a relatively stable perspective on how to address the harm. In some cases, they were no longer actively involved in the process of making sense of the harm. Recognizing that sensemaking is an ongoing and evolving endeavor that encompasses different phases \cite{weick1995sensemaking}, our recruitment criteria were designed to ensure that participants had experienced harm recently and were actively engaged in the sensemaking process.


\subsection{Post-study Survey}
After creating action plans in both conditions, participants were asked to complete a follow-up survey, where they assessed the sensemaking process from three key perspectives: the effectiveness of SnuggleSense in achieving its design objectives, the system's alignment with survivors' sensemaking goals, and the participants' ranking of SnuggleSense's individual features. A researcher was present to guide survivors through the evaluation process, prompting participants to provide rationales and posing follow-up questions after each section of the survey.

\subsubsection{How the System Meets Its Design Goals in Sensemaking}
The first part of the survey consisted of 5 rating questions. We asked participants to rate the two conditions on how well they performed along the following categories: guidance, support, agency, assistance in sensemaking, and empowerment. Participants gave a score of 1 to 7, 1 being strongly disagree and 7 being strongly agree. The first three categories, guidance, support, and agency, match the three key design guidelines of SnuggleSense: guided reflection, personalized informational support, and fostering agency through a design process. The fourth category assesses how effectively SnuggleSense achieves its primary objective of facilitating sensemaking. We included empowerment as the fifth evaluation criterion to explore how the sensemaking process might alter survivors' perceptions of their empowerment in the broader context of addressing harm.

\subsubsection{How the System Meets Survivors' Goals in Sensemaking}
Considering the varied experiences of harm that survivors have encountered, they may have distinct objectives in the sensemaking process \cite{xiao2022sensemaking}. Therefore, it is important for us to assess how participants achieve their individual goals within their specific contexts. In this step, we first asked participants to write down their goals in making sense of the harm. We then asked participants to rate how well the Unstructured and Structured processes met these goals respectively. Participants gave a score of 1 to 7, with 1 being the lowest and 7 the highest rating.

\subsubsection{Ranking of SnuggleSense Features}
Beyond assessing whether SnuggleSense achieves its intended goals, we are also interested in understanding how its various features contribute to meeting these objectives. At the end of the survey, we presented the list of features in SnuggleSense and asked participants to rank the top three that are useful to them. This approach helps to identify which elements of SnuggleSense are most instrumental in its overall effectiveness and enables participants to explain how specific features assist them.

\subsection{Initial Data Collection}
The initial collection of suggested action items in SnuggleSense was assembled from pilot testing sessions, during which participants documented action plans for addressing the harm they had encountered using SnuggleSense's individual reflection process (Figure \ref{figdesign3}, Step 1-4).  The participants in these pilot tests were selected through convenience sampling \cite{robinson_sampling_2014} of people who have experienced online interpersonal harm in the past. We obtained consent from the participants to share these action plans for experimental use. 

This initial dataset comprises contributions from 35 survivors, encompassing more than 200 action items. The action plans of pilot participants are based on a wide range of online harm experiences. 10 survivors had been called offensive names, 9 were intentionally embarrassed, 9 faced sustained harassment, 6 experienced sexual harassment, 1 was physically threatened, and 21 reported other types of harm. The incidents predominantly occurred on social media sites (31 participants), followed by texting/messaging apps (8), in person (2), personal email accounts (2), online gaming (1), forums/discussion sites (1), and online dating sites/apps (1), with 4 reports categorized as "other." In terms of the number of offenders, 14 survivors faced a single offender, 10 had 2-5 offenders, 6 had 6-10 offenders, and 5 had more than 10 offenders. The relationship with the offender varied: 17 participants were harmed by strangers, 8 by friends, and 12 by acquaintances. In Table \ref{existing_data_categories}, we presented the types of actions and stakeholders pilot participants mentioned in their action plans and their percentage in the initial dataset.

\begin{table*}
\renewcommand{\arraystretch}{0.85}
\fontsize{7}{10}\selectfont
\caption{The table shows the categories of stakeholders and actions mentioned by survivors in our initial dataset, as well as examples for each action category, collected prior to the experiment. The percentages represent the proportion of each category out of a total of over 200 action items.}
\begin{tabular}{|p{1.2cm} p{1.37cm} p{3.4cm} p{4.8cm} p{1.37cm}|}
    \hline
    Stakeholder Categories & Percentage in Initial Dataset & Action Categories & Examples & Percentage in Initial Dataset \\
    \hline
    Platform moderators & 32.58\% & Implement strategies to prevent future harm & Enforce content filters, introduce identity verification measures & 14.77\% \\
    \cline{3-5}
    & & Content moderation & Issue warnings or bans, remove offensive content & 9.09\% \\
    \cline{3-5}
    & & Give advice & Offer online harm prevention tips, share resources for managing incidents & 4.92\% \\
    \cline{3-5}
    & & Help me understand the harm & Investigate duplicate accounts, identify individuals responsible for harmful posts & 3.79\% \\
    \hline
    Offenders & 24.24\% & Understand the impact of their actions & Recognize the harm caused, understand consequences for both parties & 7.58\% \\
    \cline{3-5}
    & & Apologize & Issue a public apology, acknowledge wrongdoing & 6.44\% \\
    \cline{3-5}
    & & Explain their motivation & Disclose motivations behind harmful actions & 5.68\% \\
    \cline{3-5}
    & & Change their behavior & Commit to avoiding future harm & 3.41\% \\
    \cline{3-5}
    & & Stop the continuation of harm & Delete harmful posts & 1.14\% \\
    \hline
    Online community members & 21.21\% & Give emotional support & Reassure victims, affirm the unacceptability of online harm & 8.71\% \\
    \cline{3-5}
    & & Raise awareness & Educate about cyberbullying & 6.82\% \\
    \cline{3-5}
    & & Report inappropriate comments & Notify moderators & 3.41\% \\
    \cline{3-5}
    & & Give advice & Offer coping strategies, provide perspectives on similar experiences & 2.27\% \\
    \hline
    Family and friends & 17.05\% & Give emotional support & Offer reassurance, affirm that the victim is not at fault & 10.98\% \\
    \cline{3-5}
    & & Give advice & Suggest appropriate responses, provide guidance on handling the situation & 6.06\% \\
    \hline
    Myself & 4.92\% & Be more cautious in the future & Avoid harmful environments, be selective in online interactions & 2.27\% \\
    \cline{3-5}
    & & Communicate with offenders & Address concerns directly with the offender & 0.76\% \\
    \cline{3-5}
    & & Ignore, block, delete, leave & Disregard harmful remarks & 0.76\% \\
    \cline{3-5}
    & & Report & File a report against the offender & 0.38\% \\
    \cline{3-5}
    & & Self-care & Engage in healthy coping strategies & 0.38\% \\
    \cline{3-5}
    & & Communicate with people I trust & Consult trusted individuals for guidance & 0.38\% \\
    \hline
\end{tabular}
\label{existing_data_categories}
\end{table*}

\begin{table*}
\renewcommand{\arraystretch}{0.8}
\fontsize{7}{10}\selectfont
\newcolumntype{C}[1]{>{\arraybackslash}p{#1}}

\vspace{-3mm}
\caption{Participant demographics and experiences of online harm}
\vspace{-3mm}
\begin{tabular}{p{0.1cm} p{0.2cm} p{0.5cm} p{1cm} p{4.2cm} p{2.5cm} p{1.15cm} p{1.5cm}}
\midrule
	&	Age	&	Gender 	&	Race	&	Type of Harm	& Platform	& Number of Offender(s)	&	Relationship with Offender(s)	\\
\midrule

P1	&	19	&	Female	&	Mixed	&	Offensive name-calling, sexual harassment	&	Online dating app	&	2-5	&	Strangers	\\
P2	&	20	&	Male	&	Asian	&	Offensive name-calling, public shaming, stalking	&	Online gaming	&	6-10	&	Strangers	\\
P3	&	20	&	Female	&	White	&	Offensive name-calling	&	Social media site	&	1	&	Acquaintances	\\
P4	&	18	&	Female	&	Asian	&	Offensive name-calling, public shaming, other	&	Social media site, online gaming	&	2-5	&	Friends	\\
P5	&	22	&	Female	&	Latino	&	Public shaming	&	Social media site	&	2-5	&	Strangers	\\
P6	&	21	&	Male	&	Asian	&	Offensive name-calling, public shaming	&	Forum site	&	1	&	Strangers	\\
P7	&	20	&	Female	&	Asian	&	Offensive name-calling, physical threat	&	Social media site, forum site, in-person	&	>10	&	Strangers	\\
P8	&	19	&	Male	&	Mixed	&	Offensive name-calling, harassment, physical threat, other	&	Social media site, messaging app, in-person	&	2-5	&	Strangers	\\
P9	&	N/A	&	N/A	&	N/A	&	Stalking, other	&	Forum site	&	2-5	&	Strangers, acquaintances	\\
P10	&	19	&	Female	&	Asian	&	Offensive name-calling	&	Social media site	&	1	&	Strangers	\\
P11	&	21	&	Male	&	Asian	&	Offensive name-calling	&	Social media site	&	6-10	&	Strangers	\\
P12	&	21	&	Female	&	White	&	Other	&	Social media site	&	1	&	Strangers	\\
P13	&	20	&	Male	&	White	&	Offensive name-calling, public shaming	&	Forum site, messaging app, online gaming	&	2-5	&	Strangers	\\
P14	&	24	&	Male	&	Asian	&	Offensive name-calling	&	Messaging app	&	2-5	&	Acquaintances	\\
P15	&	19	&	Female	&	Asian	&	Public shaming, harassment	&	Social media site, in-person	&	2-5	&	Acquaintances	\\
P16	&	19	&	Female	&	Asian	&	Offensive name-calling, harassment	&	Social media site	&	>10	&	Strangers	\\
P17	&	18	&	Female	&	Asian	&	Offensive name-calling, public shaming, sexual harassment, physical threat	&	Online gaming	&	2-5	&	Strangers	\\
P18	&	20	&	Non-binary	&	White	&	Offensive name-calling, public shaming, harassment	&	Messaging app, in-person	&	2-5	&	Friends	\\
P19	&	20	&	Non-binary	&	White	&	Offensive name-calling, public shaming	&	Social media site	&	1	&	Friends	\\
P20	&	19	&	Male	&	Asian	&	Public shaming	&	Forum site	&	1	&	Strangers	\\
P21	&	21	&	Female	&	Mixed	&	Offensive name-calling, sexual harassment	&	Social media site	&	2-5	&	Strangers	\\
P22	&	18	&	Female	&	Asian	&	Offensive name-calling, public shaming, harassment, sexual harassment	&	Online gaming	&	2-5	&	Strangers	\\
P23	&	24	&	Female	&	Asian	&	Public shaming, sexual harassment	&	Social media site	&	1	&	Strangers	\\
P24	&	22	&	Female	&	Mixed	&	Offensive name-calling, harassment	&	Social media site, online dating app	&	2-5	&	Strangers	\\
P25	&	25	&	Female	&	White	&	Offensive name-calling, public shaming, other	&	Social media site	&	1	&	Strangers	\\
P26	&	18	&	Female	&	African American	&	Offensive name-calling, public shaming	&	Online gaming	&	2-5	&	Strangers	\\
P27	&	24	&	Female	&	Mixed	&	Offensive name-calling, public shaming	&	Social media site, forum site	&	1	&	Strangers	\\
P28	&	21	&	Female	&	Asian	&	Harassment, stalking	&	Social media site, in-person	&	1	&	Acquaintances	\\
P29	&	25	&	Female	&	White	&	Offensive name-calling	&	Online dating app	&	1	&	Strangers	\\
P30	&	23	&	Female	&	Asian	&	Offensive name-calling, public shaming, harassment, stalking, physical threat, other	&	Social media site, forum site, messaging app, in-person	&	2-5	&	Strangers, acquaintances	\\
P31	&	20	&	Female	&	Asian	&	Offensive name-calling, public shaming	&	Social media site	&	2-5	&	Strangers	\\
P32	&	19	&	Female	&	Mixed	&	Public shaming	&	Social media site, messaging app	&	1	&	Friends	\\
\midrule
 \end{tabular}
  \label{table:1}
\end{table*}

\subsection{Safeguarding Participants in the Experiment}
To ensure informed consent and user autonomy, users are informed of the sensemaking procedure through an introduction page before entering the system. After the sensemaking procedure, users are made aware of how their shared information will be utilized in the experiment and are given choices on whether to share their action items. Recognizing the severity of some online harm cases, we acknowledge that survivors may require additional support in addressing harm or during the sensemaking process. As a proactive measure, SnuggleSense includes a list of external resources, including non-profit organizations and a support helpline, prominently displayed at the top of the system. A researcher was present during the experiment to provide help when needed. Additionally, we mitigate the risk of exposure to inappropriate content through researchers' moderation. In both the initial dataset and any new data shared by participants while using SnuggleSense, researchers conducted a content review to ensure that it does not endorse violence or contain inappropriate material before granting access to others. These action plans are also anonymized, revealing only a general description of stakeholder types and their actions, rather than providing personally identifiable details. We used SnuggleSense in an experimental setting. As we scale the system with more survivors, we believe additional safety precautions will be necessary, which we discuss in section \ref{safeguard}.

\subsection{Participant Recruitment and Experiment Setup}
We recruited 32 participants to conduct the within-subject study from July to August 2023. We recruited these participants using a campus-wide recruiting system at a West Coast university in the United States. We randomly assigned participants to do the Unstructured condition or Structured condition first. 

We show participants' demographic information and the information about their experiences of online harm in table \ref{table:1}. Participants have an average age of 20.61, with 22 Female, 7 Male, and 2 Non-binary. There are 16 Asian, 7 White, 1 Latino, 1 African American, and 6 Mixed. One participant chose not to reveal their demographic information. Regarding their experiences with online harm, 24 participants had been called offensive names, 18 were intentionally embarrassed, 8 faced sustained harassment, 5 experienced sexual harassment, 4 were stalked, 4 were physically threatened, and 6 reported other types of harm. Most instances occurred on social media sites (20 participants), followed by forums (7), messaging apps (6), online gaming (6), and online dating apps (3). Additionally, 6 incidents had an in-person component. The number of perpetrators varied: 12 participants had a single offender, 16 had 2-5 offenders, 2 had 6-10 offenders, and 2 had more than 10 offenders. The majority (24 participants) were harmed by strangers, 6 by acquaintances, and 4 by friends. 

Participants completed the task remotely with their personal computers. A researcher was present via zoom to provide an introduction of the study in the beginning and guide the participant in the follow-up survey, and participants independently completed the task to create action plans in between. This entire process spanned approximately 50 minutes. Participants received a compensation of \$25 US dollars. The study was approved by our University's Institutional Review Board.



\section{Result}
In this section, we are looking at how well both the Unstructured and Structured conditions performed in four important areas: how the system meets its design goals, how the system meets survivors' goals in sensemaking, identifying the most useful features of SnuggleSense, and comparing the action plans in both conditions. 

\subsection{Design Goals: Guidance, Support, Agency, Sensemaking, and Empowerment}


\begin{figure*}
    \centering
    \includegraphics[width=0.8\textwidth]{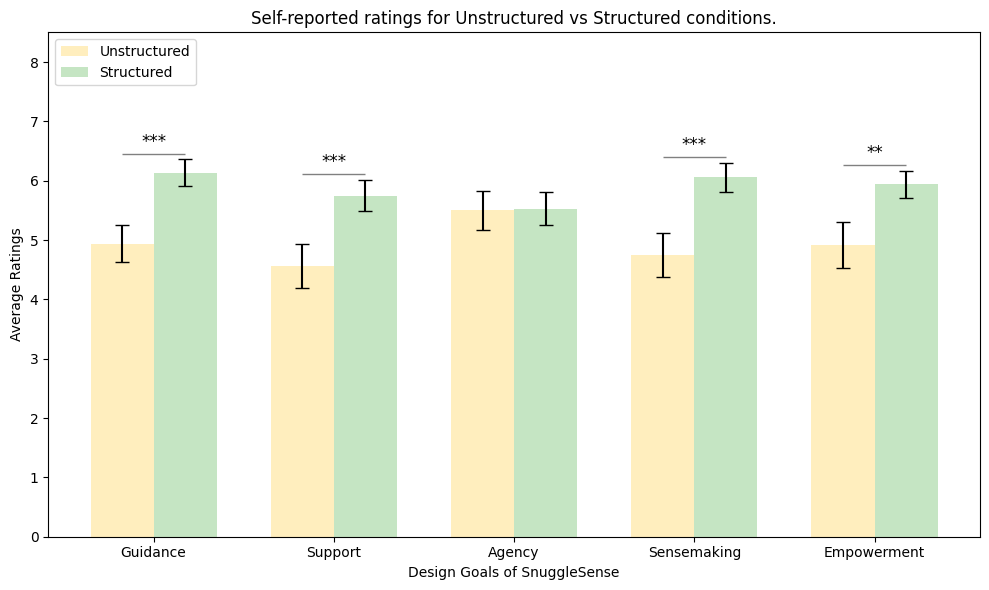}
    \caption{
    The average ratings participants gave to the 5 design goals in the Unstructured and Structured conditions. The rating scale is from 1-7, where 1 indicates strongly disagree and 7 indicates strongly agree. Two-tailed t-tests; standard deviations in parentheses, * p < .05, ** p < .01, *** p < .001. Bars signify standard error. 
    }
    \label{fig-selfreport}
    \Description{
    }
\end{figure*}

We conducted a two-tailed, paired t-test to assess the differences in ratings between the Unstructured and Structured conditions, focusing on the system’s design goals of guidance, support, agency, sensemaking, and empowerment. We presented the data in Figure \ref{fig-selfreport}. The findings indicate that the Structured condition received significantly higher ratings in guidance (Unstructured: M = 4.94, SD = 1.24; Structured: M = 6.13, SD = 0.91, p < .001), support (Unstructured: M = 4.56, SD = 1.46; Structured: M = 5.75, SD = 1.05, p < .001), sensemaking (Unstructured: M = 4.75, SD = 1.46; Structured: M = 6.06, SD = 0.98, p < .001) and empowerment (Unstructured: M = 4.91, SD = 1.55; Structured: M = 5.94, SD = 0.91, p < .01). There was no statistically significant distinction between the two conditions in terms of agency (Unstructured: M = 5.50, SD = 1.30; Structured: M = 5.53, SD = 1.11, p = n.s.). Next, we explored the rationale participants provided for their ratings across the five criteria.

\subsubsection{Guidance}
Participants favored the Structured condition for its effectiveness in providing guidance. They expressed appreciation for the step-by-step procedures: \textit{``It's very organized, very efficient in terms of guiding me to resolve the incident'' (P15)}. Moreover, participants highlighted that SnuggleSense was instrumental in breaking down complex emotional and cognitive states into manageable components, enabling them to address one problem at a time: \textit{``I think my thoughts and emotions are such a busy place…I do feel like it [SnuggleSense] just breaks it down a little more and helps me address one problem at a time and from the start to finish with the prompts" (P32).''} One participant described the sensemaking process with SnuggleSense as ``hand-holding'': \textit{``I could write down my problems as a journal and follow through very effectively...it kind of hand-holds you through the process'' (P6)}.

\subsubsection{Support}
Participants reported experiencing greater support while using SnuggleSense. The source of support most participants mentioned is the recommended action item from others who had gone through similar situations, which alleviated their feelings of isolation: \textit{``It [SnuggleSense] definitely made me feel like I wasn't alone, whereas [the Unstructured condition], it felt very like on my own and like not connected with anyone'' (P14)}. In addition, the features to facilitate sensemaking, such as the creation of sticky notes and timelines, were also cited as offering additional support, in contrast to the Unstructured condition where participants felt they were left to navigate independently: \textit{``It was a very supportive system, like the input from other people who went through the same thing and then also being able to make the timeline and sticky notes really easily…but [the Unstructured condition] didn't have any support for that. It's just, it was all on my own'' (P20)}.

\subsubsection{Agency}
Our research did not identify significant differences in the ratings of agency between the Unstructured and Structured conditions. When examining their rationales, we discovered that participants experienced agency through different pathways. In the Structured condition, agency emerged from the diverse approaches available for exploring harm and the sense of control facilitated by the design features: \textit{``There was definitely a lot more freedom with the sticky note process. With the interface of like when we had to put it on the timeline, I liked that you could actually just drag them anywhere…It felt more in my control, so I like that a lot'' (P14).} Conversely, in the Unstructured condition, agency resulted from participants owning the process themselves, free from the need to adhere to prescribed steps: \textit{``You have more freedom or like flexibility in terms of how you want to approach it [the sensemaking process]...there is more flexibility in the sense that gives you more options to practice the actual process to address the harm'' (P31)}. Some survivors preferred to navigate the sensemaking process independently rather than seeking suggestions from others: \textit{``I’d think through my own actions, more myself versus the sticky note condition kind of was looking into what others have done'' (P29)}. Participants also emphasized that it was not an either-or choice, as both conditions allowed them to retain agency over the action plan: \textit{``I thought they both provided agency because we owned the action plan in both cases'' (P29)}.

\subsubsection{Sensemaking}
Participants found the Structured condition more effective for sensemaking. Participants constantly cite the design features and the structure of SnuggleSense as helping with sensemaking, such as the timeline and the guided reflection process. The Unstructured condition, in contrast, was seen as more open-ended and less directive. To many participants, SnuggleSense provides a roadmap for thinking about the harm, not only at its occurrence but also in planning for the future:

\begin{quote}
    \textit{``The [Unstructured] one, it didn't really like, tell me how to address it. I just kind of wrote what happened to me and that was it. But with the sticky note, it was really helpful, like the whole timeline thing to actually like reorder my steps and like, see what I would do in that process…really helped just to like, delineate how I'm going to address this in the future'' (P26)}.
\end{quote}


\subsubsection{Empowerment}
The Structured condition was seen as more empowering than the Unstructured condition, especially because it allowed participants to see the thought processes of others who had faced similar issues: \textit{``A lot of the time people are hesitant to involve with authority figures because they feel like their problems aren't worth it. So to see that other people were having the same thoughts of me was empowering'' (P18)}. Some participants also expressed that agency given by the design features is empowering: \textit{``The sticky note one felt empowering because I felt that I could delete stuff or add stuff and then seeing what other people wrote and then the timeline, like being able to think through what I would want to do first and then to move forward with and having a timeline of things to do was just empowering'' (P16)}. The same participant noted that the Unstructured condition was also considered empowering, but in a way that left participants to create their own unique solutions: \textit{``[The Unstructured condition] was empowering in a different way where I created my own solutions completely and then just had some sort of framework to do something but it was in a different way, I guess'' (P16)}.

\subsection{Self-defined Goals}
\begin{table*}
 \caption{In the follow-up survey, each participant has written down a number of self-defined goals for their sensemaking of harm. The graph presents the major categories of goals participants mentioned, the percentage of participants who mentioned each category, and examples of the goals that participants created.}
\begin{tabular}{|p{4.0cm}|p{2.5cm}|p{6.0cm}|} \hline
Categories  &	Percentage of participants mentioning the category	&	Examples from participants	\\	\hline
Understand and assess the harm itself	&	40.63\%	&	Understanding why the person wanted to cause me harm, identifying who’s at fault	\\	\hline
Come up with an action plan	&	31.25\%	&	Developing structured actions for the incident, thinking about how to move forward	\\	\hline
Manage emotions or engage in self-care	&	43.75\%	&	Understanding it is not my fault, separate myself from the situation	\\	\hline
Specify actions by stakeholders (including actions by themselves)	&	43.75\%	&	Connecting with a support network in order to receive help, discussing with loved/trusted ones like family	\\	\hline
Prevent harm from happening in the future	&	28.13\%	&	Learn to respect my own boundaries, move forward and prevent similar situations from happening	\\	\hline
Actively address the harm	&	6.25\%	&    Addressing the harm that was taken, working through possible ramifications/consequences of the harm	\\	\hline
\end{tabular}
\label{self_goals}
\end{table*}

In the follow-up interview, participants were asked to establish self-defined goals for their sensemaking of harm and rate the effectiveness of the two conditions on each goal using a scale of 1 to 7, with 1 being the lowest and 7 the highest rating. Table \ref{self_goals} presents a summary of the key metrics along with the example goals articulated by participants. We conducted qualitative coding of participants' self-defined goals and the majority of them could be categorized into six categories: (1) Understand and assess the harm itself, (2) Come up with an action plan, (3) Manage emotions or engage in self-care, (4) Specify actions by stakeholders (including actions by themselves), (5) Prevent harm from happening in the future, (6) Actively address the harm. 

To assess how well the two conditions aligned with participants' self-defined goals, we conducted a two-tailed, paired t-test on the arithmetic mean ratings of the self-defined goals each participant gave to the two conditions. The findings indicate that the Structured condition received significantly higher ratings in meeting participants' self-defined goals (Unstructured: M = 4.35, SD = 1.56; Structured: M = 5.56, SD = 1.29; p < .001) \footnote{We used the arithmetic mean ratings for the t-test because of the diverse range of goals articulated by each participant. It is important to note that participants were not explicitly instructed to weigh their goals, and, therefore, the assumption of equal weighting is a limitation of this analysis.}.

\subsection{The Most Useful Features}
Our analysis identified the three features of SnuggleSense that participants found most useful: receiving recommendations (mentioned by 65.63\% of participants as top three), sorting action items on a timeline (mentioned by 59.38\% of participants as top three) and creating stakeholders and actions (mentioned by 53.13\% of participants as top three). Receiving recommendations emerged as the most frequently cited useful feature, chosen as the single most useful by 37.50\% of participants.

Next, we delved into participants' rationales for selecting the top three most useful features. In addition, we discussed how sharing action plans with others in SnuggleSense enhances a sense of connection and collaboration among its users.

\subsubsection{Receiving Recommendations}
Participants expressed that the recommended actions provided by SnuggleSense were pertinent to the challenges they were facing. As one participant explained, \textit{``Suggestions they give me, they're so tailor-made to kind of similar problems I'm dealing with and it helps inspire me to different ways to address the situation'' (P32)}. Participants appreciate the offered insights from individuals with similar experiences, which inspired them when devising their own courses of action: \textit{``I think a huge point in this thing's favor is that it shows you how other people dealt with the same issue and that gives you a lot more ideas than just trying to think on your own'' (P6)}.  

Moreover, participants highlighted the emotional validation they derived from reading about others' action items. This validation reinforced the notion that they were not alone in their struggles, as expressed by one participant: \textit{``To know that other people are also feeling the same way or similar ways as you are is very validating'' (P3)}.

\subsubsection{Sorting Action Items on a Timeline}
Participants appreciated the ability to sort their action items on a timeline, as it helped them organize their thoughts and visualize their action plans. The timeline served as a practical tool for planning actions, whether they were to be taken in response to an imminent situation or during a potential recurrence of the harm.

\begin{quote}
    \textit{``My top priority when something like this happens is usually to assess it in my brain, sort of rationally, and decide where to go from there. What's my next course of action? So sorting recommendations down on a timeline really helped me order my thoughts.'' (P17)}
\end{quote}
\begin{quote}
    \textit{``It was very useful to visualize my plan of actions that I would take, say like if this were to happen again ... I'd immediately be able to take action instead of just kind of like be in shock.'' (P26)}
\end{quote}

\subsubsection{Creating Stakeholders and Actions}
By identifying the various stakeholders involved, participants were better equipped to understand the complexity of the situation. Participants indicated that the step helps them to identify root cause of harm and alleviate their self-blame: \textit{``Not doing that [creating stakeholders and actions] makes all the pain jumble up into one, and it can cause very ineffective or not healthy ways of coping with the problem if you're not identifying what really is causing you pain'' (P7)}.

Moreover, creating stakeholders and actions allowed participants to assign responsibility for addressing the harm. Rather than merely feeling distressed about the situation, they could proactively identify individuals who could instigate change: \textit{``Instead of just like, you know, feeling bad about the situation, you can actually be like, okay, this person can actually change something, and then like thinking about what could be done is pretty helpful'' (P4)}.

\subsubsection{Sharing Action Plan with Others}
It is worth noting that while sharing their action plan with others is not an integral part of survivors’ sensemaking process of their own experiences of harm, we found that it introduced a sense of community and fostered an empowering give-and-take dynamic. Participants emphasized that sharing their action plans made them feel like they were part of a community. In contrast to the solitary act of simply writing down their thoughts, sharing created a sense of connection and contribution: \textit{``It makes you feel part of a community. And when you're just writing things down, you don't really get that, but when you share it, you kind of feel like you're contributing back and helping more people in the future'' (P6)}. This mutual exchange of knowledge and support was perceived as therapeutic: \textit{``I liked that not only was I able to use other people's recommendations, I could also submit my own for someone else that might need it in the future. So it's not just a take, it's a give and take... That's therapeutic'' (P7)}.

\subsection{Action Plan}
In this section, we examine how participants formulated their action plans in both the Unstructured and Structured conditions. We compared the types and proportions of stakeholders and actions included in participants' plans across the two conditions. Additionally, to assess how participants used SnuggleSense’s recommendations, we analyzed the types and proportions of recommended action items that participants incorporated into their plans.

\subsubsection{Number of Stakeholders and Actions}
We counted the number of distinct stakeholders and action items participants mentioned in the action plan of both conditions. The data revealed that participants incorporated significantly more stakeholders in their action plan when using the Structured condition (M = 4.34, SD = 1.64) compared to the Unstructured condition (M = 3.16, SD = 1.08), p < .001. In addition, participants formulated significantly more action items in the Structured condition (M = 6.25, SD = 2.87) than the Unstructured condition (M = 4.50, SD = 2.23), p < .001.

\subsubsection{Categories of Stakeholders and Actions}
Table \ref{stakeholder_category} and Table \ref{action_category} present the main stakeholder and action categories mentioned by participants in both the Unstructured and Structured conditions. These tables also provide the percentage of participants who referenced each category in each condition and the percentage point difference between the two conditions.

In the Structured condition, participants displayed a greater tendency to involve various stakeholder types (e.g., family and friends, platform moderators, and online community members) rather than assigning actions to themselves (as shown in Table \ref{stakeholder_category}).
When analyzing the shift in action types from the Unstructured to the Structured condition, we see a reduction in instances of self-directed problem solving, such as opting for actions like “Ignore, block, delete, leave” (refer to Table \ref{action_category}). Simultaneously, there is a significant increase in participants seeking explanations from offenders (a 28.13\% increase) and soliciting advice (a 34.38\% increase) or emotional support (a 28.13\% increase) from online community members.

The data suggests that the Structured condition leads participants to consider a more diverse and inclusive range of stakeholders, shifting focus from a self-centric approach in the Unstructured condition to a more community- and network-centric perspective. It also promotes the consideration of approaches beyond content moderation as the sole means of addressing harm, encouraging actions that delve into the underlying causes of harm and seek support from various sources. This shift is echoed in the reflections of participants:

\begin{quote}
    \textit{``There's something in [the Structured condition] that you see, it [the harm] is a systemic problem. It's not just some random one bad guy in the world that wants to send harmful messages. There could be a lot of other people involved that can, you know, make this issue better.'' (P2)}
\end{quote}

\begin{quote}
    \textit{``When I was just writing things down [in the Unstructured condition], I wasn't thinking as much about who was at fault, but I think [the Structured condition] helped to clarify that a bit more and just understand that it wasn't really on me for what happened.'' (P13)}
\end{quote}

\begin{table*}
 \caption{Comparison of Stakeholder Types Mentioned by Participants in Unstructured and Structured Conditions (ordered by percentage point difference). The first column presents the primary stakeholder categories identified by participants in both conditions. The second and third columns depict the percentage of participants who mentioned the stakeholder category in each condition. The final column illustrates the percentage point difference between the two conditions.}
\begin{tabular}{|p{4.0cm}|p{1.9cm}|p{1.6cm}|p{4.6cm}|} \hline
Stakeholder Categories	&	Unstructured	&	Structured	&	Percentage point difference (Structured minus Unstructured)	\\	\hline
Myself	&	68.75\%	&	53.13\%	&	-15.62\%	\\	\hline
Offenders	&	75\%	&	81.25\%	&	6.25\%	\\	\hline
Platform moderators	&	78.13\%	&	90.63\%	&	12.50\%	\\	\hline
Online community members	&	25\%	&	53.13\%	&	28.13\%	\\	\hline
Family and friends	&	37.50\%	&	68.75\%	&	31.25\%	\\	\hline
\end{tabular}
\label{stakeholder_category}
\end{table*}

\begin{table*}
 \caption{Comparison of Action Categories Mentioned by Participants in Unstructured and Structured Conditions (ordered by percentage point difference). The first and second columns present the primary stakeholder and action categories identified by participants in both conditions. The third and fourth columns depict the percentage of participants who mentioned the action category in each condition. The final column illustrates the percentage point difference between the two conditions.}
\begin{tabular}{|p{2.9cm}|p{3.5cm}|p{1.8cm}|p{1.5cm}|p{2.4cm}|} \hline
Stakeholder Categories	&	Action Categories	&	Unstructured	&	Structured	&	Percentage point difference (Structured minus Unstructured)	\\	\hline
Myself	&	Ignore, block, delete, leave	&	53.13\%	&	28.13\%	&	-25.00\%	\\	\hline
Family and friends	&	Give emotional support	&	28.13\%	&	40.63\%	&	12.50\%	\\	\hline
Offenders	&	Stop the continuation of harm	&	31.25\%	&	43.75\%	&	12.50\%	\\	\hline
Platform moderators	&	Content moderation	&	72\%	&	87.50\%	&	15.63\%	\\	\hline
Family and friends	&	Give advice	&	19\%	&	40.63\%	&	21.88\%	\\	\hline
Offenders	&	Explain their motivation	&	6.25\%	&	34.38\%	&	28.13\%	\\	\hline
Online community members	&	Give emotional support	&	9.38\%	&	37.50\%	&	28.13\%	\\	\hline
Online community members	&	Give advice	&	6.25\%	&	40.63\%	&	34.38\%	\\	\hline
\end{tabular}
\label{action_category}
\end{table*}

\subsubsection{Adopted Suggestions}
In the Structured condition, a majority of participants (81.25\%) adopted SnuggleSense's recommended action items into their own action plans, and 42.22\% of participants' action items were derived from SnuggleSense's recommendations. This aligns with participants' identification of recommended action items as the most valuable feature.

Furthermore, 65.63\% of participants added new categories of stakeholders they had not previously considered. Specifically, 40.63\% added online community members as a new category, 25\% added themselves, 15.63\% added friends and family, 15.63\% added platform moderators, and 12.50\% added the offender.

Additionally, 81.25\% of participants added new actions to existing stakeholders or new stakeholders. The top three new categories of actions participants added were for platform moderators to implement strategies to prevent future harm (40.63\%), for online community members to give emotional support (28.13\%), and for offenders to explain their motivations for conducting harm (25\%).
\section{Discussion}
In this paper, we introduced SnuggleSense, a system designed to empower survivors through a structured sensemaking process. Our evaluation demonstrates its effectiveness in enhancing survivors' sensemaking. In this section, we reflect on these findings and explore their implications for addressing online harm and offering support to survivors in online spaces with social computing systems. We first argue that the sensemaking process enabled by SnuggleSense has the potential to empower online harm survivors, granting them agency and power in meeting their needs to address harm. Next, we explore how SnuggleSense opens up a restorative justice pathway for harm resolution. In the end, we reflect on our design lessons, future work, and limitations.

\subsection{Sensemaking as a Process towards Survivor Empowerment}
Empowerment is the process by which individuals and collectives gain control over issues that affect them \cite{rappaport1987terms, fawcett1994contextual}. In the context of online harm, the empowerment of survivors can be seen as the process where survivors gain control over how to address the harm they experience. 
In the following, we discuss how the sensemaking process in SnuggleSense empowers survivors from two perspectives: first, by providing a structure for sensemaking, and second, by making survivors aware of their support communities and the resources available to them.

\subsubsection{Empowerment through a structured sensemaking process}
The structured sensemaking of SnuggleSense enables survivors to establish a clearer and more actionable connection between their goals in addressing harm and the means to achieve them. Currently, survivors often need to develop an action plan for addressing harm on their own or turn to online support groups, where they share their experiences and seek advice and support from the community in a question-and-answer format \cite{barta2023similar, andalibi2017sensitive, 10.1145/3617654}. Our research found that using SnuggleSense provides significantly more guidance than simply coming up with actions to address the harm, and sorting actions on a timeline is one of the most important features perceived by participants. Zimmerman argues that empowerment occurs when individuals can perceive a direct correspondence between their goals and how to attain them \cite{zimmerman2000empowerment}. The structured sensemaking process of SnuggleSense, with its guided reflective questions and the visualization of action plans on the timeline, helps survivors gain a deeper understanding of their experiences and how to address their needs.

In addition, the structured sensemaking process in SnuggleSense enhances the knowledge survivors need to address the harm they experience. Our results show that participants using SnuggleSense developed action plans with a wider range of actions and stakeholders. A key aspect of psychological empowerment is increasing awareness of available actions and strengthening problem-solving skills \cite{schneider2018empowerment}. Through its guided reflection questions and recommended actions, SnuggleSense provides a framework for survivors to reflect on their experiences and needs, expanding their understanding of potential actions to address the harm they face.

\subsubsection{Empowerment through awareness of communities and resources}
Sense of empowerment can be enhanced by sense of community \cite{chavis1990sense}, as well as the ability to identify those with power, resources, and connections to the issue of  \cite{zimmerman2000empowerment}. While social media platforms often present content moderation or blocking as the only available options for survivors \cite{roberts2019behind,gillespie2018custodians}, SnuggleSense empowers survivors by fostering awareness of the communities they are part of and the support and resources available to them. The community SnuggleSense introduces are two-fold. First, it encourages survivors to consider their social circles, such as family and friends, or the online communities where the harm occurred. Participants highly valued the function of identifying stakeholders and their actions in SnuggleSense, mentioning significantly more stakeholders and their actions compared to the Unstructured condition in their plans. 

Second, SnuggleSense facilitates survivors to find inspiration and validation in other survivors' experiences. Survivors rated receiving recommendations as the top useful feature. Further, survivors are also empowered by sharing and contributing to other survivors who use SnuggleSense. Participants highly valued the ability to share action plans and inspire others. It gives them a sense of community and they gain agency and control through giving back to the community. Zimmerman believes that being involved in community organizations can exercise a sense of competence and control \cite{zimmerman2000empowerment}. Survivors derive strength from one another, and the willingness to share their plans with others demonstrates the platform's potential to foster a sense of community among survivors.

An empowered individual is essential for empowered communities \cite{laverack2006improving, zimmerman2000empowerment}. In addition, connecting with more stakeholders facilitates community empowerment by raising awareness of a problem’s existence and negotiating common goal \cite{li2018working}. Besides aiding survivors in addressing current harm, we envision SnuggleSense as a tool that also serves to educate and empower the community in the long run. SnuggleSense offers a sensemaking framework that can be applied to future instances of harm experienced by a survivor or others.

\subsection{A Restorative Justice Approach to Addressing Online Harm}
Our results also indicate how a restorative justice pathway empowers survivors to consider community-based harm resolutions and prioritize restoration and healing. Our research indicates a shift in survivors' responses when utilizing SnuggleSense, involving a broader array of online and offline stakeholders, including family, friends, and online community members, in the process of addressing harm. In addition, survivors move away from individual efforts such as blocking, muting, or solely relying on punitive measures of moderators to understand the motivations behind harm or seek emotional support.

These observed shifts align closely with the recommendations put forth by the researcher community, emphasizing the need for designing interventions that prioritize survivors' healing and restoration needs \cite{schoenebeck2021drawing, musgrave2022experiences, xiao2022sensemaking, goyal2022you, sultana2022shishushurokkha}. SnuggleSense builds on this foundation by integrating restorative justice practices into online spaces, a practice inherently designed to support these needs. Moreover, our findings resonate with the work of researchers who embrace a community-based approach to addressing harm. For instance, Squadbox employs ``friend-sourcing'' to empower survivors \cite{maharSquadboxToolCombat2018}, while Heartmob relies on online community members to provide assistance \cite{blackwell_classification_2017}. SnuggleSense joins these community-based approaches by enabling survivors to find inspiration and validation from others with similar experiences.

Importantly, our research underscores the potential of restorative justice principles in achieving these transformative shifts. 
Restorative justice encourages people to identify the root cause of harm and emphasizes support and healing instead of punishing the perpetrators \cite{zehr2015little}. It locates harm in communities and argues that community members have a stake in addressing the harm \cite{zehr2015little}. It is worth noting that SnuggleSense did not explicitly dictate the stakeholders or actions involved; rather, these results emerge organically through the empowerment of survivors and their agency in the sensemaking process. SnuggleSense joins other work and shows how restorative justice provides a potential pathway in online harm-resolution that complements the current approach \cite{schoenebeck2021drawing, kou2021punishment, xiao2022sensemaking}.

\subsection{Design Insights and Future Work}
SnuggleSense demonstrates how a social computing system can support online harm survivors in the sensemaking process. Our experiments with 32 participants highlight SnuggleSense's potential to scale and assist a broader range of survivors. Moreover, we believe SnuggleSense’s design provides valuable insights for developing future social computing systems that support survivors, particularly by facilitating sensemaking and promoting community awareness. In this section, we reflect on the design lessons learned from deploying SnuggleSense in an experimental setting, with the goal of informing future work in this area.

\subsubsection{Tailored Support to Survivors}
In SnuggleSense, we provide survivors with informational support by suggesting relevant stakeholders and actions. This is achieved through algorithms that assess the similarity of survivors' responses to multiple-choice questions about their harm experiences. Future systems have the potential to further refine these recommendation mechanisms to better tailor support to survivors.

The similarity between survivors can be measured using diverse metrics. Recent research found that online harm survivors' needs can be influenced by various factors, including personal traits (e.g., demographics \cite{schoenebeck2021drawing, schoenebeckOnlineHarassmentMajority2023}, role in society and culture \cite{warfordSoKFrameworkUnifying2021}), past experiences with harm \cite{schoenebeck2021drawing}, or the context of harm (e.g., their relationship with the perpetrators \cite{warfordSoKFrameworkUnifying2021}, the time span of harm \cite{thomasItCommonPart2022}). These factors present opportunities for tailoring suggestions to survivors. In addition, these aspects may influence survivors' needs differently and hold varying degrees of importance for different individuals. In future work, we plan to conduct large-scale surveys to explore how participants harm experiences and their personal traits influence their needs differently. Additionally, future systems can be designed to support specific populations vulnerable to online harm to provide suggestions that are better tailored to their experiences and needs.

When providing personalized recommendations, it’s important to balance guidance with agency. We acknowledge that while providing guidance can empower survivors, it can also limit their agency. Our research revealed no significant difference in how the Structured and Unstructured conditions provide a sense of agency to survivors. When participants explained their preferences, some found the Structured condition offered more freedom and control by allowing them to take ownership of the design process. In contrast, others appreciated the Unstructured condition as it required them to think more deeply about their actions without external guidance. In SnuggleSense, we chose to let survivors initially reflect on the harm independently before providing suggestions. Finding the optimal balance between these two objectives is an important challenge to explore in future work.

\subsubsection{Nurturing a Support Community among Survivors}
SnuggleSense highlights the potential to foster mutual aid communities among survivors of online harm. In traditional online support groups, help often comes from bystanders or community members who may not share the survivors' experiences. Prior research has explored how individuals seek support on social media platforms, such as using Reddit throwaway accounts \cite{ammari2019self} or engaging with the \#Depression tag on Instagram \cite{andalibi2017sensitive}. Platforms such as Heartmob allow those experiencing online harassment to share their stories, receive supportive messages, or request help in reporting harassers from bystanders \cite{blackwell_classification_2017}. 
However, these approaches encounter challenges. Survivors may experience secondary harm from individuals who lack a deep understanding of their experiences \cite{to2020they}. Furthermore, differing perspectives on how to address harm—often from bystanders or external stakeholders—may not align with survivors' actual needs or desires \cite{xiao2023addressing}.

SnuggleSense offers an alternative by highlighting the essential role survivors can play in addressing harm within their own community. Through mutual exchanges, survivors share contextually relevant advice, affirmation, and validation. This aligns with Fraser’s work on the value of self-paced, internal discussions among marginalized groups \cite{fraserRethinkingPublicSphere1990}. By creating a space for survivors to share their experiences and action plans, SnuggleSense empowers individuals to explore and affirm unique strategies for addressing harm—strategies that are often overlooked by traditional content moderation systems.

SnuggleSense invites us to explore the potential of creating systems that foster survivor-led support communities. SnuggleSense facilitates the asynchronous exchange of action plans, enabling survivors to find informational support even when external resources are unavailable. Our findings show that participants value this reciprocal dynamic: receiving suggested actions was the most appreciated feature, and survivors felt a sense of reward for contributing to the community. Thus, future systems can consider supporting more varied forms of interaction within survivor communities. Survivors could validate others' proposed actions, share insights, or even return to the platform to provide updates on their progress after addressing harm. By fostering a cycle of support, such systems have the potential to nurture a supportive network that leads to community empowerment.

\subsubsection{Safeguarding Survivors} \label{safeguard}
Participants in our study used SnuggleSense in an experimental setting. Scaling systems that support survivors requires additional safety and privacy considerations. Survivors face risks of secondary harm from other online members or their perpetrators \cite{to2020they, xiao2023addressing}. Like existing online support communities, content moderation will be necessary to identify and remove inappropriate content \cite{blackwell_classification_2017, ammari2019self, andalibi2017sensitive}. 

To protect survivors' privacy, the system should provide clear information about how their data is stored and shared, giving them control over whether to delete or withhold it. It can also incorporate guidelines encouraging participants to avoid sharing personally identifiable information when discussing their experiences. Furthermore, to better safeguard survivors—particularly those with prior experiences of harm—the system’s design can be guided by trauma-informed principles \cite{chen2022trauma, scott2023trauma}.

Survivors' experiences and needs are individualized and may change over time \cite{xiao2022sensemaking, weick1995sensemaking}. When using algorithms to provide personalized suggestions, it is important to assess how these algorithms influence survivors' decision-making processes and whether they deliver recommendations that cater to survivors' needs and the system’s goals \cite{yurrita2022towards,karusala2024understanding, saxena2024algorithmic}. It should also continuously review and update security and safety measures, ensuring survivors can modify or revoke their consent as their needs evolve \cite{im2021yes}.

\subsubsection{Beyond Sensemaking: Taking Actions}
While SnuggleSense focuses on the sensemaking stage, taking action is a crucial component in achieving empowerment in practice \cite{zimmerman2000empowerment, musgrave2022experiences, blackwell_classification_2017}. There are challenges for survivors to implement the actions they propose. SnuggleSense provides avenues for addressing harm that are not traditionally applied, making it hard for survivors to envision the effectiveness of those alternatives \cite{xiao2023addressing}. Therefore, the motivation to act on the action plans can be directly linked to the availability and accessibility of resources for survivors to act on their newfound understanding. We believe that it is essential to pair the improvement of survivors' sensemaking process through SunggleSense with the allocation of resources and the creation of supportive conditions for survivors to act. This includes creating ways to assemble relevant stakeholders and resources to assist survivors \cite{sultana_unmochon_2021, maharSquadboxToolCombat2018, blackwell_classification_2017, goyal2022you}, or changing societal attitudes toward addressing harm \cite{xiao2023addressing}. In future research, we aim to conduct longitudinal studies to explore how survivors continue to engage with their action plans developed through SnuggleSense, with a focus on identifying, designing, and consolidating resources to support survivors in the ongoing process of addressing harm. We also plan to develop tools to support stakeholders, such as moderators, perpetrators, and community members, to collectively make sense of and address harm.




\subsection{Limitations}
We studied SnuggleSense in an experimental setting. Applying these results to commercial-sized platforms and broader use cases would require adapting the design to suit the specific demands and complexities of those contexts.

Our pilot and study participants primarily comprised college students in the United States, potentially limiting the generalizability of our findings to other survivor demographics. Additionally, the harm scenarios shared by participants may not encompass the full spectrum of online harm experiences. Our initial dataset contains over 200 action items, with survivor similarity calculated based on a limited set of harm experience dimensions. This may limit the diversity of recommendations provided to participants. 

Sensemaking is a dynamic and evolving process, influenced by various factors over time \cite{weick1995sensemaking}. Our study imposed a finite timeframe for participants to make sense of a given harm scenario. It is plausible that participants' perceptions of the same incident might undergo changes with extended time for sensemaking. Furthermore, it is imperative to recognize that sensemaking represents an initial step towards addressing harm. The subsequent action taken is integral to empowerment of survivors \cite{zimmerman2000empowerment}. Therefore, a comprehensive assessment of our system's effectiveness can only be achieved through evaluating its impact in the later stages of executing the action plan, which constitutes an avenue for our future research.

\section{Conclusion}
Our paper introduces SnuggleSense, a system designed to empower survivors of online harm by guiding them through a sensemaking process. Inspired by restorative justice, SnuggleSense opens up new opportunities for survivors to assert their agency and define their paths toward healing and resolution. SnuggleSense represents a step forward in empowering survivors of online harm centering their needs and agency in the sensemaking process and highlighting the importance of providing them with the tools, support, and community-based resources to address harm.

\begin{acks}
We thank the anonymous reviewers and our participants for their valuable feedback and contributions.
\end{acks}

\bibliographystyle{ACM-Reference-Format}
\bibliography{references,s-references,My_Library}

\end{document}